**Reducing the strain required for ambient-pressure superconductivity in bilayer nickelates**

*Yaoju Tarn[1,2,#,\*], Yidi Liu[2,3,#], Florian Theuss[2,5], Jiarui Li[2], Bai Yang Wang[2,3], Jiayue Wang[1,2,5], Vivek Thampy[4], Zhi-Xun Shen[1,2,3,5], Yijun Yu[1,2,6,\*], Harold Y. Hwang[1,2,\*].*

[1] Department of Applied Physics, Stanford University, Stanford CA, USA

[2] Stanford Institute for Materials and Energy Sciences, SLAC National Accelerator Laboratory, Menlo Park CA, USA

[3] Department of Physics, Stanford University, Stanford CA, USA

[4] Stanford Synchrotron Radiation Lightsource, SLAC National Accelerator Laboratory, Menlo Park CA, USA

[5] Geballe Laboratory for Advanced Materials, Department of Physics and Applied Physics, Stanford University, Stanford CA, USA

[6] Department of Physics, Fudan University, Shanghai, China

[#] These authors contributed equally

Corresponding E-mail: ytarn@stanford.edu, yuyijun@fudan.edu.cn, hyhwang@stanford.edu

**Abstract:** The remarkable discovery of high temperature superconductivity in bulk bilayer nickelates under high pressure has prompted the conjecture that epitaxial compressive strain might mimic essential aspects of hydrostatic pressure. The successful realization of superconductivity in films on $SrLaAlO_4$ (001) (SLAO) supports this correspondence, yet it remains unclear whether the rich pressure–temperature phase diagram of bilayer nickelates can be systematically mapped (and studied at ambient pressure) as a function of epitaxial strain. To this end, experimental access near the elusive edge of the superconducting phase boundary would provide invaluable insight into the nature of the superconducting state and the ground state from which it emerges. It would also offer a benchmark for theoretical models. Here we report superconducting bilayer nickelates grown on $LaAlO_3$ (001) (LAO), where the compressive strain required for ambient-pressure superconductivity is nearly halved to -1.2%. These films exhibit a superconducting onset above 10 K and reach zero resistance at 3 K, with normal-state transport properties differing from those of films grown on SLAO. Our results offer a new opportunity to probe emergent phenomena near the superconducting phase boundary in the strain–temperature phase diagram of bilayer nickelates.

## 1. Introduction

The discovery of superconductivity under hydrostatic pressure in bilayer nickelates establishes a new family of transition-metal-oxide superconductors characterized by a $d^{7.5}$ electronic configuration.[1–6] Inspired by the concept of mimicking hydrostatic pressure using in-plane biaxial epitaxial strain, bilayer nickelate thin films grown on compressively straining SLAO substrates exhibit ambient-pressure superconductivity, with an onset transition temperature ($T_{c,onset} \approx 40$–$50$ K)[7–10], roughly half of the bulk maximum (80–90 K) under pressure. While the application of pressure to epitaxial thin films on other substrates also induces superconductivity in this system,[11] ambient-pressure superconductivity has thus far only been realized on SLAO, which imposes a large compressive strain of about –2.0%. Nevertheless, this sole instance invites a number of intriguing questions: What is the minimum strain required to sustain superconductivity? How faithfully can the rich pressure-temperature bulk phase diagram be mapped onto a strain-temperature phase diagram in thin films? What ground state emerges when superconductivity is suppressed? This is a particularly compelling question as, historically, crucial insights into superconductivity have often arisen from its neighboring phases.[12]

From a practical standpoint, exploring superconductivity under weaker compressive strain is also appealing, as (i) the large lattice mismatch on SLAO (-2.0%) constrains the coherent growth to ultrathin films below 10 nm.[13] Reduced strain, on the other hand, is expected to allow the realization of thicker and higher quality films[14] and a cleaner manifestation of superconductivity and neighboring phases; (ii) a high superconducting transition temperature ($T_c$) obscures the normal-state properties, making studies like normal-state resistivity[15] or quantum oscillations[16,17] possible only under extreme magnetic fields. Similarly, spectroscopic measurements that probe features near the Fermi surfaces are either dominated by superconducting features below $T_c$, or severe thermal broadening above $T_c$.

In this work, we stabilize superconductivity in $La_2PrNi_2O_7$ (LPNO) thin films on LAO substrates, thereby pushing the lower bound of compressive strain required for superconductivity in bilayer nickelate films from -2% to -1.2%. As expected, $T_c$ decreases, but we also observe notable contrasts in normal-state behavior compared to bilayer nickelates grown on SLAO. This advance broadens the experimental window for exploring superconducting bilayer nickelate thin films, addressing some of the limitations in samples grown on SLAO mentioned above. Furthermore, it offers access near the boundary of the superconducting phase, which may provide valuable insight into the nature of the superconducting state and the proximate ground state. From here, we refer to superconducting bilayer nickelates LPNO grown on LAO (SLAO) as LPNO/LAO (LPNO/SLAO), respectively.

## 2. Results & Discussion

Our initial efforts began with the growth of $La_3Ni_2O_7$ films on LAO by molecular beam epitaxy (MBE)[18] following the discovery of superconductivity in $La_3Ni_2O_7$ films grown on SLAO.[7] A signature of a partial superconducting transition was observed after post-growth ozone annealing (**Figure S1**). Motivated by subsequent improvements achieved through Pr substitution in both bulk and thin films,[5,8,9] we then shifted our focus to LPNO. LPNO thin films (7-10 nm extracted from Scherrer fitting and Laue oscillations) were synthesized on LAO and capped with a protective layer of 1–2-unit cells of $SrTiO_3$ (STO) via pulsed laser deposition (PLD) (see **Methods**). We observe a wider range of growth conditions (oxygen pressure and growth temperature) than that of LPNO/SLAO grown in the same chamber[9] (**Figure S2(a)**), and generally find coherent growth is more straightforward on LAO than on SLAO, presumably due to the lower strain mismatch (**Figure S2(b)**). Following the growth, we applied an ex-situ ozone annealing process with resistivity feedback[9] to induce superconductivity by filling oxygen vacancies.[19] Representative resistivity evolution profiles during ozone annealing are shown in **Figure S3**.

We characterized the superconductivity of two typical LPNO/LAO samples by measuring their resistivity as a function of temperature $\rho(T)$ and comparing them with representative LPNO/SLAO in Ref. [9] (**Figure 1(a)**). Our LPNO samples, labelled A and B, exhibit an $T_{c,onset}$ (defined graphically in the right inset of **Figure 1(a)**) above 10 K and a zero-resistance $T_c$ ($T_{c,zero}$) above 3 K (the left inset of **Figure 1(a)**). The diamagnetic response of sample B was also observed via two-coil mutual inductance[20] (MI, **Figure 1(b)**). The onset of the MI signal occurs at 4 K, in agreement with the $T_{c,zero}$ obtained in $\rho(T)$. The extracted London penetration depth, assuming a Scherrer film thickness of 8 nm, is roughly $\lambda(0) = 4$ μm, consistent with literature on LPNO/SLAO.[8] Next, we measured critical current density ($J_c$) of sample B and obtain its electric field ($E$) versus current density ($J$) characteristics from 1.5 to 5.0 K (**Figure 1(c)**). We extract $J_c$ as a function of temperature (**Figure 1(d)**) and find that $J_c$ drops to zero at around 3.6 K, in agreement with the value of $T_{c,zero}$ inferred from $\rho(T)$ and MI measurements. $J_c$ at 1.5 K is around 1.2 kA $cm^{-2}$, about ten times lower than in LPNO/SLAO,[9] consistent with the observation that the highest $T_{c,zero}$ observed in LPNO/LAO is also ten times smaller.

Next, we study the response of superconductivity to magnetic fields parallel and perpendicular to the $c$-axis (**Figure 2(a)** and **(b),** respectively) in LPNO/LAO by measuring $\rho(T)$ under different fields in sample B. We determine the upper critical fields ($H_{c2}$) in each orientation by fitting to the linearized Ginzburg-Landau (GL) model (**Equations 1, 2**), which is applicable for a superconductor that is geometrically constrained in thickness,

$$H_{c2,\parallel c} = \frac{\phi_0}{2\pi\,\xi_{ab}^2(0)}\left(1 - \frac{T}{T_c}\right) \tag{1}$$

$$H_{c2,\parallel ab} = \frac{\sqrt{12}\phi_0}{2\pi\,\xi_{ab}(0)d}\sqrt{1 - \frac{T}{T_c}} \tag{2}$$

where $\phi_0$ is the superconducting flux quantum, $\xi_{ab}(0)$ is the in-plane GL coherence length, and $d$ is the superconducting thickness. We obtain $H_{c2,//c}$ = 12.4 T and $H_{c2,//ab}$ = 19.5 T, respectively, smaller than or close to the Pauli limit (18 T). An in-plane GL coherence length of $\xi_{ab}(0)$ = 5.2 nm and a superconducting thickness of $d$ = 11 nm are also obtained (**Figure 3(c)**, slightly larger than the 8 nm Scherrer thickness and 9 nm obtained by Laue oscillations), indicating the bulk nature of the superconductivity. The extracted $H_{c2}$ is about five to six times smaller than in LPNO/SLAO,[8–10,15] while the anisotropy factor $\gamma$=1.6 remains comparable.[8,9,15] This $\gamma$ value is significantly lower than that of typical two-dimensional or layered superconductors,[21] further supporting the bulk nature of the superconductivity in LPNO on both substrates. While these lower $H_{c2}$ values are within expectations, as the $T_{c,onset}$ and $T_{c,zero}$ between the two systems also scales by approximately this factor (10 K and 3 K on LAO versus 50 K and 30 K on SLAO respectively), they make LPNO/LAO an experimentally accessible platform to study the ground state after superconductivity is suppressed by field. In LPNO/SLAO, the high $H_{c2}$ (>50 T) makes it prohibitively difficult to access the normal-state properties at low temperatures.[15] In contrast, LPNO/LAO offers a convenient platform for such studies and could yield valuable, otherwise inaccessible insights into bilayer nickelates, thereby complementing the electronic structure information obtainable by surface-sensitive probes.[18,22–24]

To gain insights on the structural properties of superconducting bilayer-nickelate thin films, we measured $\theta$–$2\theta$ symmetric scan and reciprocal space mapping (RSM) by X-ray diffraction (XRD) of LPNO/LAO. We observe strong Bragg peaks from the LPNO film along with clear Laue oscillations in the $\theta$–$2\theta$ symmetric scan (**Figure 3(a)**) of sample A, indicating high crystalline quality and film uniformity. The RSM of a superconducting LPNO/LAO sample shows that the LAO (10$\bar{3}$) and LPNO (10$\underline{17}$) peak both lie on the $H = 1$ line, indicating that the film is coherently strained to the substrate (**Figure 3(b)**), with the corresponding pseudo-cubic in-plane lattice constant $a_p = 3.787$ Å). Comparing the $T_{c,onset}$ of LPNO/LAO and its lattice constants with those of other bilayer nickelates in the literature across pressurized bulk[1,5,6] (empty black symbols) and epitaxial thin films[7–10,24,25] (colored filled symbols) at ambient pressure (**Figure 3(c)**), we find that LPNO/LAO follows the common trend of $T_{c,onset}$ versus $a_p$ shared by both bulk and SLAO-grown thin films. Meanwhile, as a function of out-of-plane lattice constant $c$ (**Figure 3(d)**), there is a contrast in behavior between thin films and bulk crystals. LPNO/LAO and previously reported (La, Pr, Sm, Sr)$_3$Ni$_2$O$_7$ [7–10,24,25] exhibit a roughly positively correlated relationship between $T_{c,onset}$ and $c$, opposite to bulk studies. Unlike in bulk crystals, where $a_p$ and $c$ both contract under pressure, epitaxial thin films respond differently: compressive (tensile) strain reduces $a_p$ while elongating (contracting) $c$, following the film's Poisson ratio. It is worth noting that scanning transmission electron microscopy (STEM) studies show bilayer nickelates on LAO have similar Ni-planar O bond tilting pattern as in superconducting bilayer nickelates on SLAO or under pressure.[26]

LPNO/LAO now represents the superconducting member closest to a presumed superconducting phase boundary in the strain–temperature phase diagram (**Figure 3(c)**). With limited knowledge of this boundary from both thin films and bulk, its nature remains uncertain: it could represent either a quantum critical point, or a first-order transition (e.g. structural transition as in Ref. [26]) without criticality. In other words, superconductivity might emerge continuously from zero temperature as compressive strain increases, or instead appear abruptly at a finite transition temperature that is truncated by the first-order boundary. To this end, analysis of the normal-state transport properties of LPNO/LAO may provide some insights. Indeed, we observe an intriguing contrast between LPNO/LAO and LPNO/SLAO. First, we consider that resistivity of correlated metals is often expressed either as $\rho \sim \alpha_2 T^2 + \alpha_1 T + \rho_0$ (where $\alpha_2$ and $\alpha_1$ are the $T$-quadratic and $T$-linear coefficients, and $\rho_0$ is the residual resistivity),[27,28] or in the power law form $\rho \sim AT^n + \rho_0$ (where $n$ is the exponent and $A$ the pre-factor).[29] We find that neither form yields a satisfactory fit across the entire normal-state regime of LPNO/SLAO and LPNO/LAO, as shown in **Figure S4**. To obtain a reliable fit, it is necessary to include a Mott-Ioffe-Regel (MIR) saturation resistivity term, $\rho_{MIR}$,[30] following the so-called parallel-resistor formalism (PRF), as expressed in **Equation 3** and **4**

$$\frac{1}{\rho(T)} = \frac{1}{\rho_0 + AT^n} + \frac{1}{\rho_{MIR}} \tag{3}$$

$$\frac{1}{\rho(T)} = \frac{1}{\rho_0 + \alpha_1 T + \alpha_2 T^2} + \frac{1}{\rho_{MIR}} \tag{4}$$

The PRF in both forms has been used to fit the normal-state resistivity of cuprates across the superconducting dome,[31,32] rare-earth perovskite nickelates across a large range of strain states,[33] and more recently in LPNO/SLAO.[9,10,15] We fit the $\rho(T)$ curves of six LPNO/SLAO samples in literature[8,9,34] and six LPNO/LAO samples from $T_{c,onset}$ + 5 K to the highest available temperature in the data, by both **Equation 3** and **4** as shown in **Figure 4** and **Figure S5**, respectively. Both equations yield equally satisfactory fits. In the following, we focus on the results obtained using **Equation 3**, while those based on **Equation 4**, presented in the **Supplementary Information**, yield consistent conclusions.

In **Equation 3**, in the case of the exponent $n$ = 2, the prefactor $A$ typically reflects electron-electron interactions and the system exhibits Fermi liquid behavior.[35] The case of $1 \leq n < 2$ has been interpreted as being in a non-Fermi-liquid (NFL) or marginal Fermi-liquid (MFL)[36,37] regime. We find a robust difference in $n$ but similar $\rho_{MIR}$ (**Table 1**): The exponent in LPNO/SLAO is $n$ = 2 within error as expected, but is approximately $n$ = 5/3 in LPNO/LAO (**Figure 4(c)**). $\rho_{MIR}$ values, on the other hand, overlap between the two substrates (with some scatter; **Figure 4(d)**). $\rho_{MIR}$ for both substrates averages around 0.9 mΩ cm, with a lower bound of 0.6 mΩ cm, corresponding to the 0.65 mΩ cm $\rho_{MIR}$ estimated from Ref. [38] after considering the different numbers of NiO$_2$ planes per thickness. $n < 2$ behavior in resistivity is typically associated either with electron–phonon interactions[39] or linked to quantum critical fluctuations.[40] In the former scenario, electron-phonon interactions can often be stronger in more compressively strained systems, opposite to what we observe here. The effective strength of this coupling in bilayer nickelates may, however, be modulated by their specific atomic structure and phonon spectra under strain conditions. In the latter scenario, the emergence of NFL behavior in the less-compressed films may reflect enhanced quantum critical scattering as the system approaches a putative quantum critical point. The $n$ = 5/3 power law is characteristic of scattering near a three-dimensional ferromagnetic quantum critical point.[29] Nonetheless, such a mechanism would be rather unexpected in the bilayer nickelates. Whether alternative scattering processes (e.g. possible spin fluctuations) can give rise to a similar power law remains to be explored. Notably, a similar crossover from $n$ = 2 to $n$ = 5/3 has also been observed in perovskite nickelates under strain[33] and pressure,[41] suggesting a possible shared underlying mechanism among nickelates. Comparison of the normal-state transport properties in bilayer nickelates across different strain states thus reveals a landscape of features, warranting further investigation to elucidate its microscopic origin and its potential connection to superconductivity.

## 3. Conclusion

The realization of ambient-pressure superconductivity in bilayer nickelates grown on LAO nearly halves the epitaxial compressive strain required to stablize superconductivity in thin films, opening an avenue to study the system near the superconducting phase boundary. Additionally, it provides a new platform with a different strain state to compare hydrostatic pressure with epitaxial compression, potentially elucidating the common structural aspects of superconductivity. The five-fold reduction in $H_{c2}$ facilitates access to the normal state at low temperature, which is inaccessible in films grown on SLAO. The reduced lattice mismatch relative to SLAO makes synthesis more straightforward and enables thicker superconducting films than on SLAO, which can provide a more favorable platform for spectroscopy and other signal-limited experiments. Altogether, the realization of ambient-pressure superconductivity in LPNO/LAO provides another experimentally valuable and theoretically rich component for studies of superconductivity in the nickelates.

## 4. Methods

*Sample preparation*

As-received LAO (001) substrates from Shinkosha were cleaned in acetone and IPA before growth. The substrate was heated to 680°C as measured by a pyrometer. The oxygen partial pressure during the growth was maintained at 150 mTorr. The LPNO film and STO cap were grown by ablating a polycrystalline LPNO target and single-crystal STO substrate respectively. The laser fluence was 0.56 J cm$^{-2}$ with a 1.8 x 1.8 mm spot size. The repetition rate for growing the film (STO cap) was 5 Hz (3 Hz). After growth, the temperature was reduced to below 560°C and held for 2 minutes and then quench cooled. We find empirically that this process reduces the amount of visible twinning caused by the unavoidable cubic-to-rhombohedral structural phase transition[42] as LAO cools.

*Structural characterization*

Standard $\theta$-$2\theta$ symmetric scans were conducted on a lab-based X-ray diffractometer immediately after growth using a wavelength of $\lambda = 1.5406$ Å. High-resolution scans and RSM were conducted at the Stanford Synchrotron Radiation Lightsource (SSRL) on B17-2 using a Pilatus 3 100K-M detector with photon energy 8.333 keV. The data in **Figure 3(a)** was an L-scan taken with this energy and converted to angle using $\lambda$ = 1.5406 Å. Out-of-plane lattice constants are determined by quoting a reported value in cited work if available, or by the position of the (0012) film peak if not. Scans in the supplementary information were conducted on a Rigaku SmartLab.

*Ozone annealing process*

We use the same bespoke ozone anneal setup as detailed in Ref. [9] The optimized ozone anneal condition is similar to that of LPNO/SLAO samples. We monitor in-situ resistivity of the films throughout the ozone annealing process and stop at resistivity saturation. We similarly find that a combination of high temperature and low ozone density is required to maximize the filling of oxygen vacancies while minimizing the transformation to higher-order RP phases. Representative examples are shown in **Figure S2**. We warm up and cool down rapidly to quickly cross the phase boundaries between nearby RP phases as discussed in Ref. [9].

*Transport and mutual inductance measurements*

As-grown samples were cut and deposited with 40 nm of gold by electron-beam evaporation to make contact electrodes defined by shadow masks in either van der Pauw or Hall bar geometry, then bonded to ceramic chip carriers by ultrasonic wire bonding before performing ozone annealing. Low-temperature transport measurements were conducted in helium cryostats. Two-coil mutual inductance measurements were performed in a dilution refrigerator with drive coil current of 50 mA.[20] We show the Hall coefficient of samples A and B in **Figure S6**.

**Author Contributions**

Y.T. grew the samples under the mentorship of Y.L., with B.Y.W. and Z.-X.S. identifying the initial signatures of superconductivity. Y.T. performed the ozone annealing and transport measurements with assistance from Y.L. and Y.Y. F.T. and Y.T. carried out the mutual inductance measurements. J.L., J.W., Y.T., Y.L., and V.T conducted the structural characterization. Y.Y. and H.Y.H. supervised the project. Y.T., Y.Y., and H.Y.H. analyzed the data and wrote the manuscript with input from all authors.

**Data Availability Statement**

Source data is available upon reasonable requests to the first and/or corresponding authors.

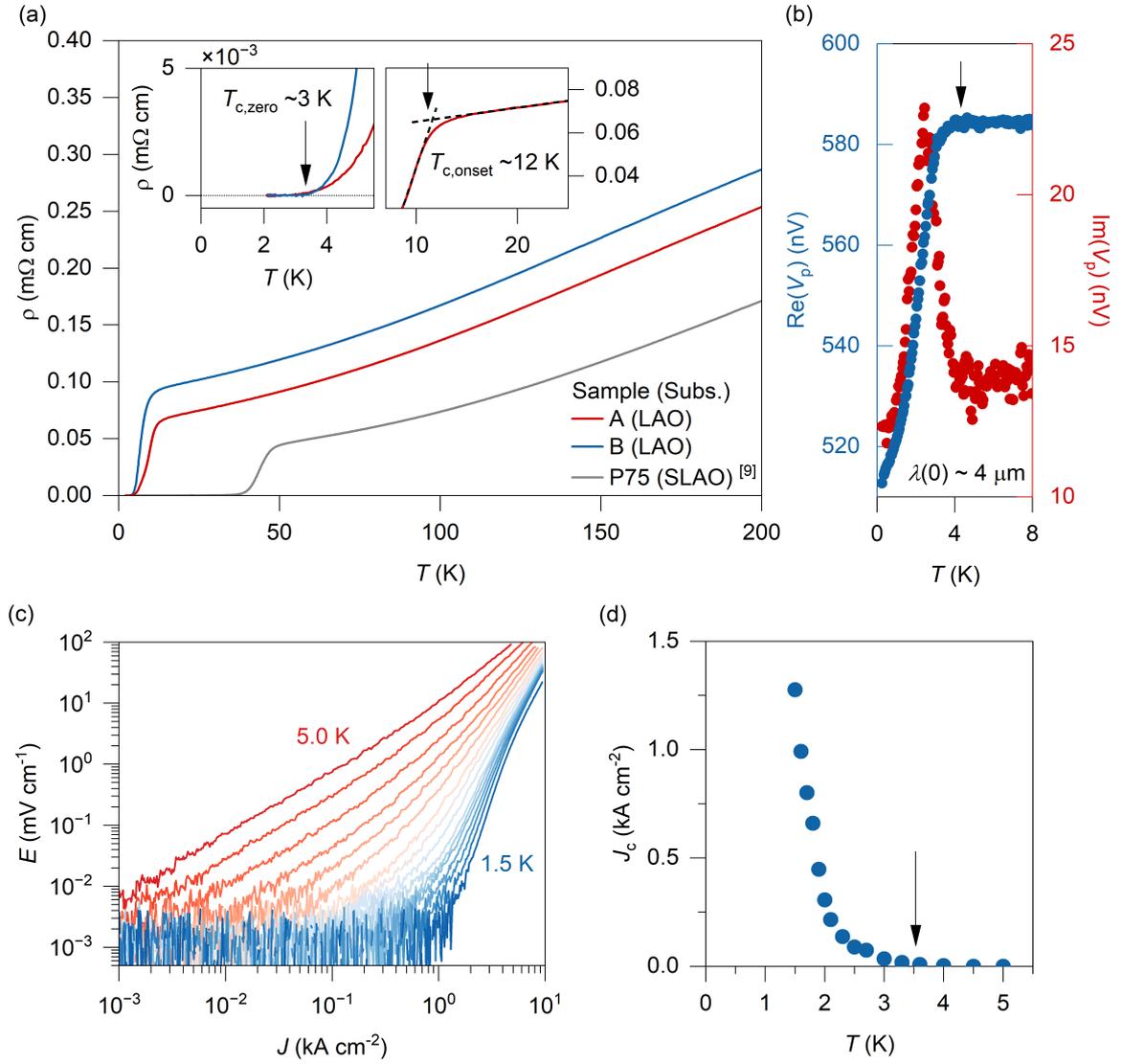

**Figure 1. Ambient pressure superconductivity in LPNO thin films grown on LAO. (a)** $\rho(T)$ for thin films of LPNO grown on substrates LAO and SLAO (from Ref. [9]). Insets indicate zero resistance ($T_{c,zero}$) and the onset of superconductivity ($T_{c,onset}$). For films grown on LAO, the $T_{c,zero}$ and $T_{c,onset}$ are around 3 K and 12 K respectively. **(b)** Diamagnetic response of LPNO/LAO sample B probed by mutual inductance. The real (imaginary) part of the response function is plotted on the left (right) axis. The arrow indicates the approximate onset of the diamagnetic response. **(c)** $E$-$J$ characteristics of sample B. From the right to left, the data are taken from 1.5 K to 2.1 K in steps of 0.1 K, and 2.3, 2.5, 2.7, 3.0, 3.3, 3.6, 4.0, 4.5, 5.0 K. **(d)** $J_c$ as a function of $T$, as extracted from **(c)**. The black arrow indicates $T_{c,zero}$. $J_c$ is defined as the value of $J$ when $E$ is higher than the noise floor of electric field limited by our measurement electronics, which corresponds to 3 μV cm⁻¹.

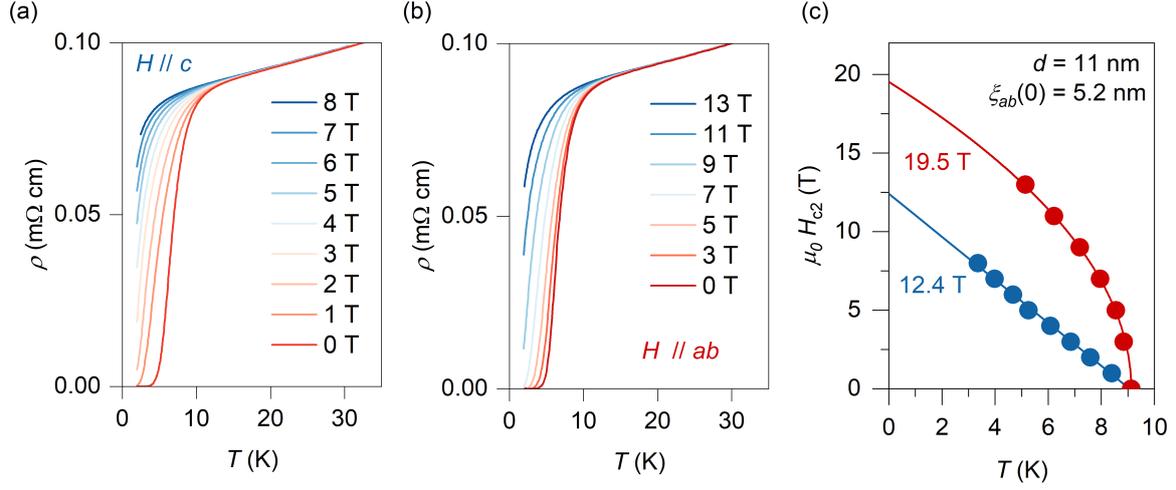

**Figure 2. Superconductivity in external magnetic fields. (a)** and **(b)** Plots of $\rho(T)$ of sample B under various applied magnetic fields aligned parallel to the $c$-axis and $ab$-plane of the sample, respectively. **(c)** Extracted $H_{c2}$ values for both field orientations, and their corresponding Ginzburg-Landau fits. At zero temperature, $H_{c2,//c}$ = 12.4 T, $H_{c2,//ab}$ = 19.5 T. The corresponding in-plane coherence length $\xi_{ab}(0)$ = 5.2 nm and the superconducting thickness $d$ = 11 nm. All references to $H_{c2}$ use the criterion of the temperature when resistivity crosses 90% of the zero-field resistivity at $T_{c,\text{onset}}$.

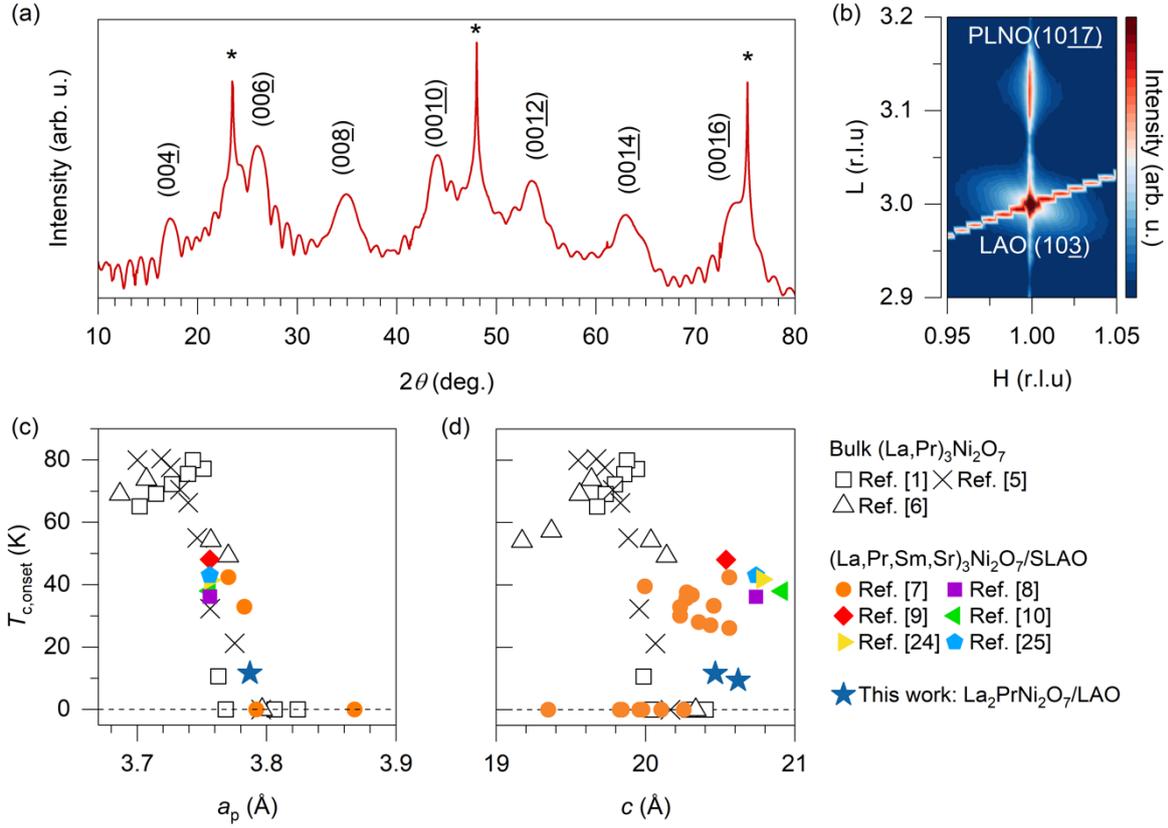

**Figure 3. Structural characterization of superconducting films. (a)** High-resolution $\theta$–$2\theta$ symmetric scan of a superconducting LPNO film grown on LAO (sample A). The asterisks (*) mark LAO substrate peaks (001), (002), and (003) from left to right. The LPNO film peaks are labelled (00$\underline{L}$). **(b)** Reciprocal space map (RSM) of a superconducting LPNO film. Axes are in reciprocal lattice units (r.l.u.). **(c, d)** Comparison of $T_{c,onset}$ versus in-plane (out-of-plane) lattice constant $a_p$ ($c$). The empty black symbols are from literature on bulk bilayer nickelates, while the colored symbols are from literature on thin films at ambient pressure. The points in panel **(c)** are extracted from the highest $T_{c,onset}$ samples from each reference, using the definition of $T_{c,onset}$ as in **Figure 1(a)** inset. The legend is shared between **(c)** and **(d)**.

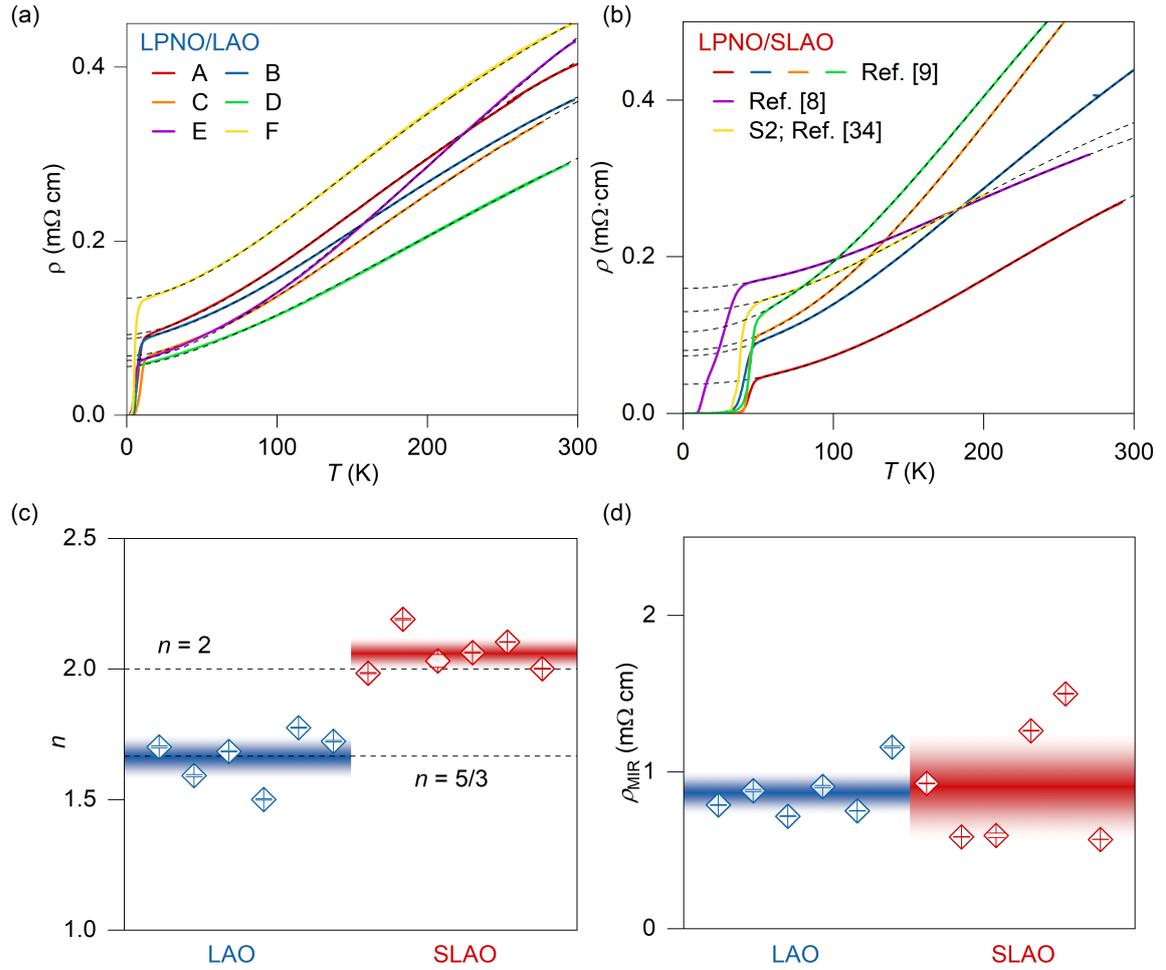

**Figure 4. Comparison of normal state resistivity in LPNO/LAO and LPNO/SLAO. (a-b)** $\rho(T)$ of six LPNO/LAO samples **(a)** and six LPNO/SLAO samples **(b)** reported in the literature (Refs. [8,9,34]) fitted with **Equation 3**, respectively. **(c-d)** Comparison of $n$ and $\rho_{MIR}$ for LPNO/LAO (blue) and LPNO/SLAO (red), respectively. All error bars represent the standard deviations of the parameters obtained from the fitting. The shaded horizontal bands indicate the average values of each parameter, with their widths corresponding to twice the standard deviation derived from statistics over six samples.

|  | $n$ | $\rho_{\text{MIR}}$ (m$\Omega$ cm) |
|---|---|---|
| LPNO/SLAO | $2.06 \pm 0.07$ | $0.91 \pm 0.39$ |
| LPNO/LAO | $1.66 \pm 0.09$ | $0.87 \pm 0.16$ |

**Table 1:** Average values of fit parameters $n$ and $\rho_{\text{MIR}}$ for LPNO/SLAO and LPNO/LAO. Values are averages over six samples and uncertainties are the standard deviations.

# Supporting Information

**Supporting Information for "Reducing the strain required for ambient-pressure superconductivity in bilayer nickelates"**

*Yaoju Tarn[1,2,#,*], Yidi Liu[2,3,#], Florian Theuss[2,5], Jiarui Li[2], Bai Yang Wang[2,3], Jiayue Wang[1,2,5], Vivek Thampy[4], Zhi-Xun Shen[1,2,3,5], Yijun Yu[1,2,6,*], Harold Y. Hwang[1,2,*].*

[1] Department of Applied Physics, Stanford University, Stanford CA, USA

[2] Stanford Institute for Materials and Energy Sciences, SLAC National Accelerator Laboratory, Menlo Park CA, USA

[3] Department of Physics, Stanford University, Stanford CA, USA

[4] Stanford Synchrotron Radiation Lightsource, SLAC National Accelerator Laboratory, Menlo Park CA, USA

[5] Geballe Laboratory for Advanced Materials, Department of Physics and Applied Physics, Stanford University, Stanford CA, USA

[6] Department of Physics, Fudan University, Shanghai, China

[#] These authors contributed equally

Corresponding E-mail: ytarn@stanford.edu, yuyijun@fudan.edu.cn, hyhwang@stanford.edu

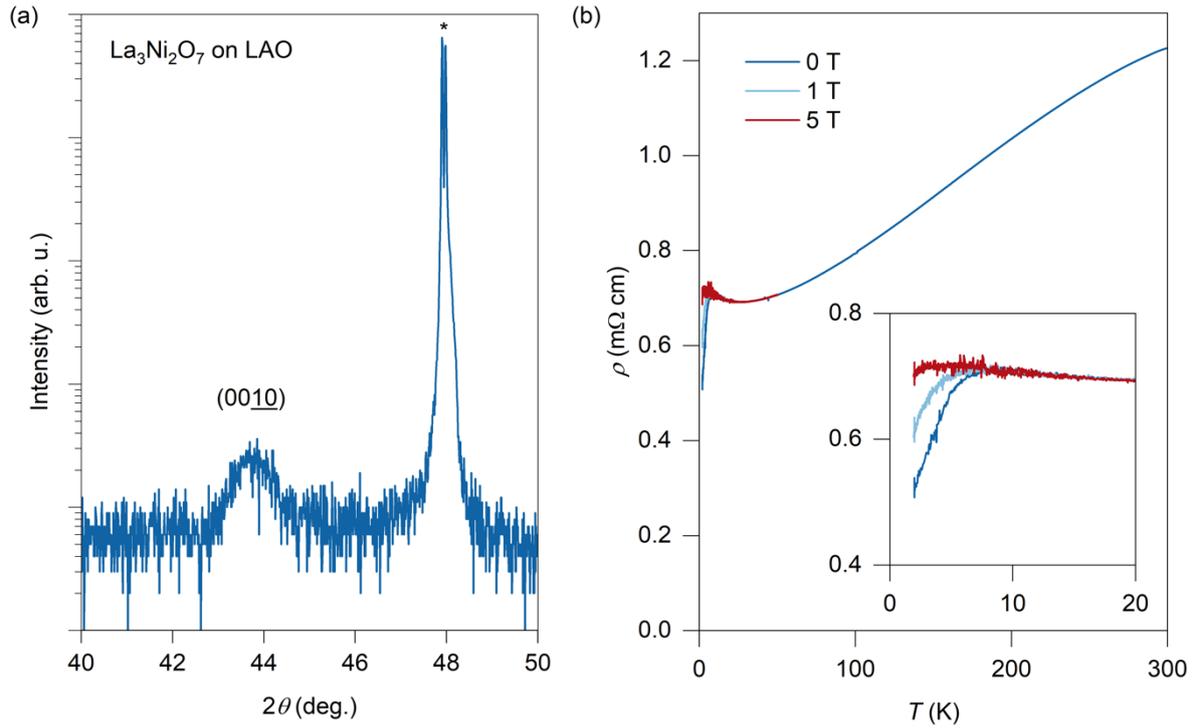

**Figure S1. Initial signatures of superconductivity of La$_3$Ni$_2$O$_7$ films on LAO**. **(a)** A La$_3$Ni$_2$O$_7$ sample grown by MBE on LAO characterized by a $\theta$–$2\theta$ symmetric XRD scan. The substrate peak is marked with an asterisk (*) **(b)** $\rho(T)$ of the sample shown in **(a)** after ozone annealing, at an applied magnetic field of 0 T, 1 T, and 5 T. The inset shows an expanded view near $T_{\text{c,onset}}$.

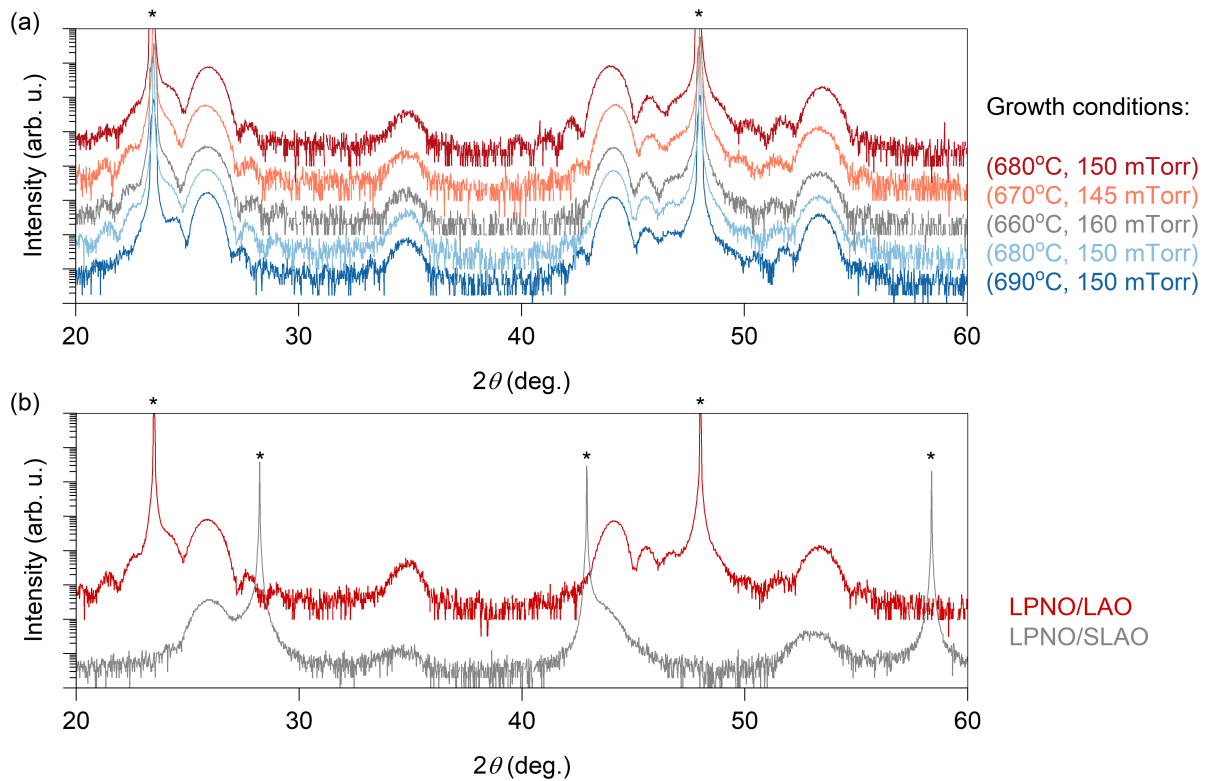

**Figure S2. Relative stability of growth conditions.** **(a)** As-grown LPNO/LAO samples characterized by $\theta$–$2\theta$ symmetric XRD scans. Our nominally optimized growth condition is at a substrate surface temperature of 680°C and at an oxygen pressure of 150 mTorr. We vary our growth condition through a range of temperatures and pressures and find that our growth process is stable within a 30°C and 15 mTorr window. Films grown in this range show largely consistent and reproducible results in XRD. This window of stability appears to be less challenging than that of LPNO/SLAO (Ref. [9], main text), presumably due to reduced strain. The growth conditions for the samples shown here vary across the range of (substrate temperature, oxygen partial pressure) = (660-690 °C, 145-160 mTorr), as color-coded for each XRD scan. **(b)** Comparison of LPNO grown on LAO (red) and SLAO (grey) in the same chamber measured on the same XRD system. The LPNO/LAO sample overall exhibits higher crystallinity. All substrate peaks are marked with asterisks (*).

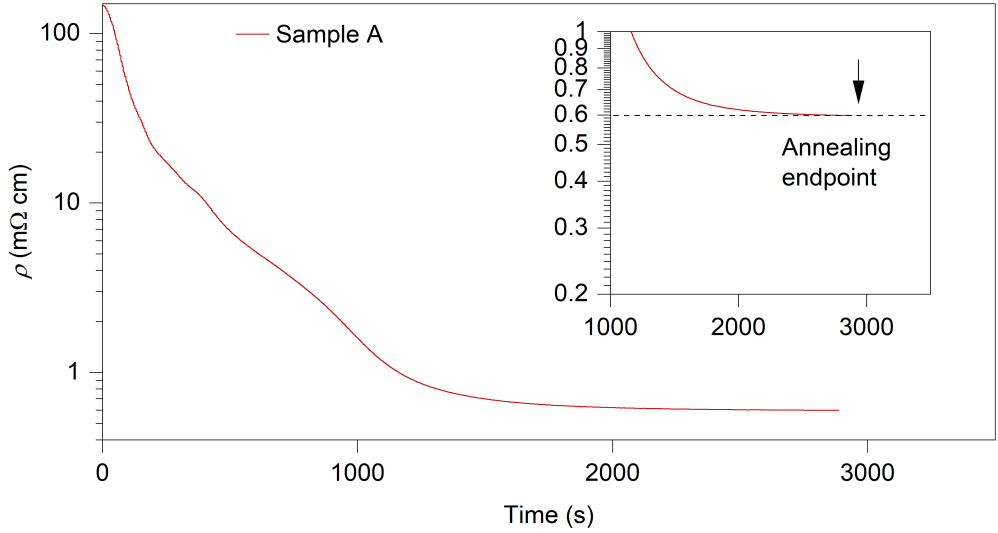

**Figure S3. Representative ozone annealing profiles for superconducting LPNO/LAO samples.** The resistivity, tracked in-situ in a tube furnace, of a piece of sample A as it was annealed in ozone at 305°C from its as-grown state. The warm-up and cool down times are limited to 5 and 10 minutes respectively. The arrow indicates where a minimum resistivity appears to occur, after which the annealing process is stopped by quickly removing the tube from the furnace. We find empirically that optimal samples will saturate under a resistivity of 1 mΩ cm.

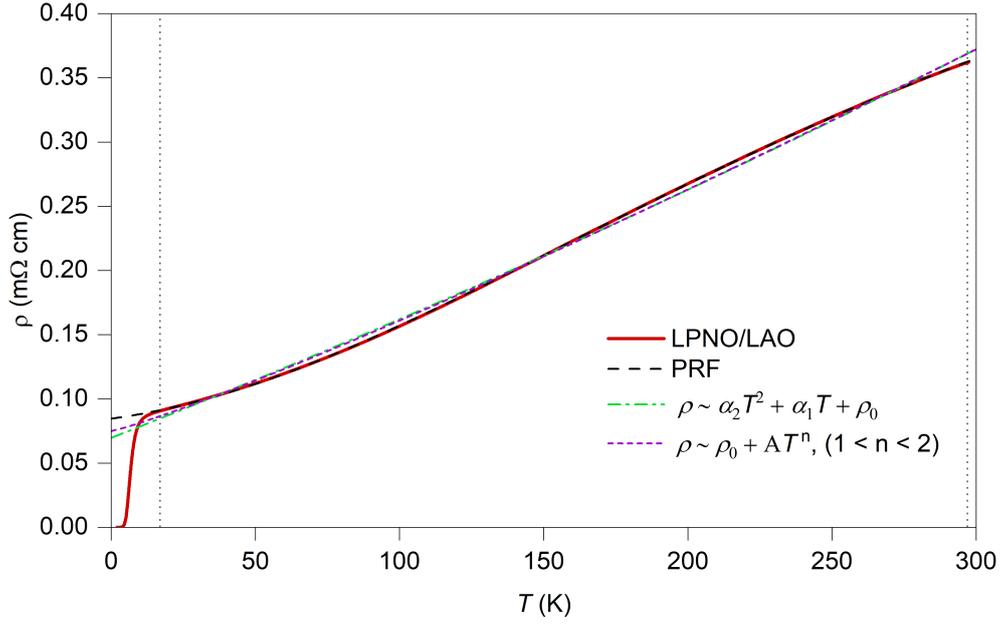

**Figure S4. Comparison of functional forms for normal state fitting without a $\rho_{MIR}$ term.** The $\rho(T)$ curve of a typical superconducting LPNO/LAO sample (red) is fit from 5 K above $T_{c,onset}$ (left vertical dotted line) to around 297 K (right vertical dotted line) using several common functional forms used for normal state analysis. The PRF fit (black dashed line) tracks the entire normal state well, but neither the form $\rho \sim \alpha_2 T^2 + \alpha_1 T + \alpha_0$ (green) nor the power law fit $\rho \sim \rho_0 + AT^n$ (purple) yields a satisfactory fit.

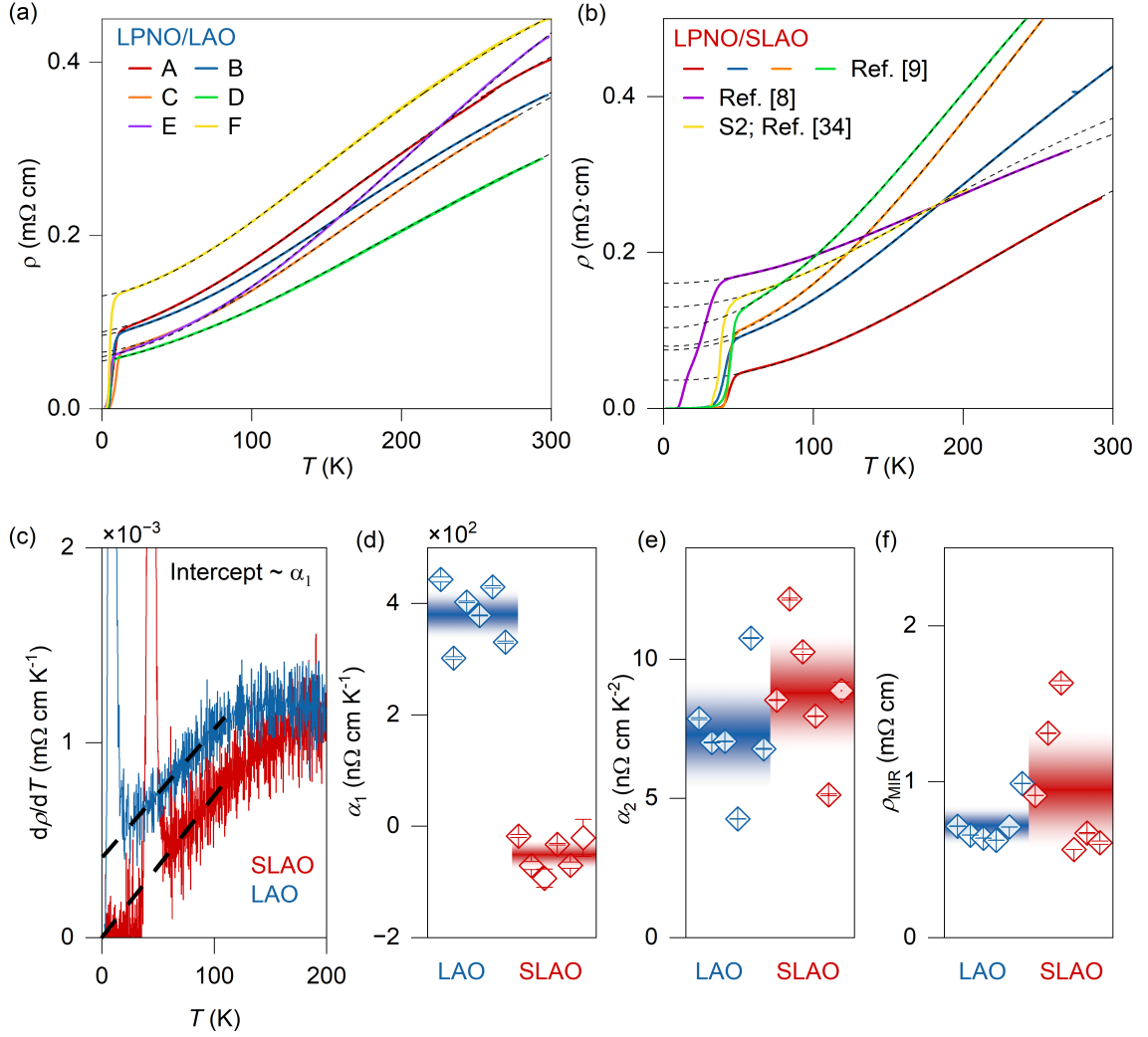

**Figure S5. Comparison of normal state resistivity in LPNO/LAO and LPNO/SLAO fitted with Equation 4. (a-b)** $\rho(T)$ of six LPNO/LAO samples **(a)** and six LPNO/SLAO samples **(b) reported** in the literature (Refs. [8,9,34]) fitted with **Equation 4**, respectively. **(c)** Comparison of $d\rho/dT$ of a sample on LAO (blue) and SLAO (red). The intercept is non-zero (near-zero) for LPNO/LAO (LPNO/SLAO). The dashed lines are guides to the eye for the $T$-linear contribution. **(d-f)** Comparison of $\alpha_1$, $\alpha_2$, and $\rho_{MIR}$ for LPNO/LAO (blue) and LPNO/SLAO (red), respectively. The average values are shown in **Table S1**. All error bars represent the standard deviations obtained from the fitting. The shaded horizontal bands indicate the average values of each parameter, with their widths corresponding to twice the standard deviation derived from statistics over six samples. The pronounced $T$-linear contribution observed in LPNO/LAO in comparison with LPNO/SLAO is consistent with the fitting results obtained using **Equation 3**, and all analyses presented in the main text remain applicable here.

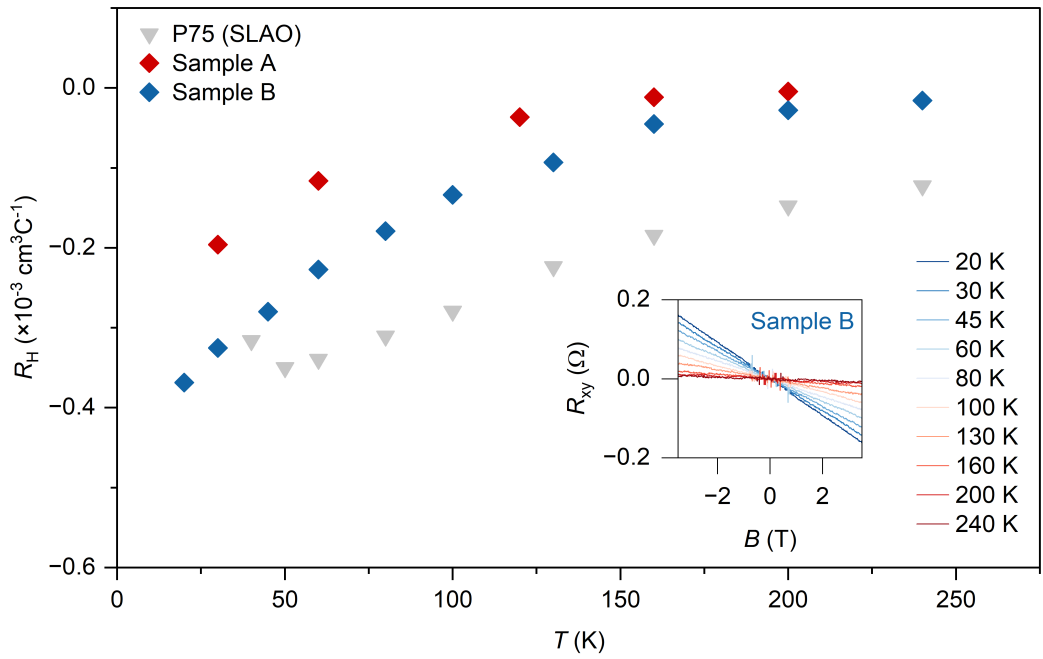

**Figure S6. Hall coefficient of superconducting LPNO/LAO.** Hall coefficient $R_H$ of superconducting LPNO/LAO samples A (red) and B (blue) over a range of temperatures. The grey triangles are from P75, a LPNO/SLAO sample from Ref. [9] in the main text. The inset shows the anti-symmetrized Hall resistance of sample B from which $R_H$ were extracted.

|  | $\alpha_1$ (nΩ cm K$^{-1}$) | $\alpha_2$ (nΩ cm K$^{-2}$) | $\rho_{MIR}$ (mΩ cm) |
|---|---|---|---|
| LPNO/SLAO | -50 ± 30 | 8.8 ± 2.4 | 0.95 ± 0.44 |
| LPNO/LAO | 380 ± 50 | 7.3 ± 2.2 | 0.72 ± 0.14 |

**Table S1:** Average values of fit parameters $\alpha_1$, $\alpha_2$, and $\rho_{MIR}$ for LPNO/SLAO and LPNO/LAO. All the errors quoted here are standard deviation derived from statistics over six samples.